\documentclass[10pt,a4paper]{article}
\usepackage{latexsym,amssymb,amsmath} 
\usepackage{citesort} 
\usepackage[]{graphicx} 
\usepackage{setspace}
\usepackage{color}
\usepackage{courier}
\definecolor{mygray}{rgb}{0.8,0.8,0.8}
\usepackage{listings}
\usepackage{algorithm}
\usepackage[noend]{algorithmic}
\usepackage{array}
\begin{document}
%
\title{Bufferless NOC Simulation of Large Multicore System on GPU Hardware}
\author{Navin Kumar, Aryabartta Sahu \ \\
Deptt. of Comp. Sc. and Engg., Indian  Institute of Technology Guwahati \\
Guwahati, Assam, India, 781039, Email: \{navin.k, asahu\}@iitg.ac.in
}
\date{}

\maketitle
\begin{abstract}
\normalsize
\textit{
Last level cache management and core interconnection network  play important roles in performance and power consumption in multicore system. Large scale chip multicore uses mesh interconnect widely due to scalability and simplicity of the mesh interconnection design. As interconnection network occupied significant area and consumes significant percent of system power, bufferless network is an appealing alternative design to reduce power consumption and hardware cost. We have designed and implemented a simulator for simulation of distributed cache management of large chip multicore where cores are connected using bufferless interconnection network. Also, we have redesigned and implemented the our simulator which is a GPU compatible parallel version of the same simulator using CUDA programming model. We have simulated target large chip multicore with up to 43,000 cores and achieved up to 25 times speedup on NVIDIA GeForce GTX 690 GPU over serial simulation. 
}
\end{abstract}

\section{Introduction}

In large scale chip multicore (LCMP), on-chip cache management and interconnection network have significant  impact  on  performance, power consumption of the system.  As the core count of chip multicore increase, the pressure on on-chip cache (in particularly the last level cache (LLC) L2 cache) increase significantly. Single shared cache (physically shared) is not good for performance in terms of access latency and interference among cores. Any how, there are many level of caches in this kind of system, first level cache (L0 an L1) must be private, but the last level cache (the L2 cache) which must be bigger and need to be managed efficiently.  Also the completely distributed (physically distributed) cache may not be good for many cases where a core requires a larger portion of cache. Distributed cache suffers from increased local cache pressure and eviction. So logically shared and physically distributed model (LSPD) capture both performance in terms of access time and share effectively. Among various last level cache management models, LSPD model of last level cache is promising in terms of cache utilization and overall system performance. The performance of LSPD model depends on effective policy for the cache block placement, eviction, migration and directory management.

Mesh interconnection network to connect the cores in LCMP is widely used as it provides a good trade off between simplicity, scalability and maintainability. As stated in \cite{one,two,three}, the interconnection network in LCMP occupies significant amount of area and consumes around 40\% of total power, so bufferless network is promising alternative design to reduce hardware cost and power consumption where overall network traffic is low to medium range. 

It is good idea to explore all the design spaces of different cache management policies, which suite to large multicore system connected using bufferless interconnection network. So in this work, we have designed an efficient simulator to simulate on chip cache management of LCMP where cores are connected using bufferless network. As most of available simulators (targeted to simulate large scale multicore architecture) are very slow, and also our designed one is very slow, so we have parallelized our designed simulator and ported onto GPU platform using CUDA programming \cite{sixteen,seventeen} to speed up the simulation. In terms of cost of the system, a personal computer with an additional GPU card is cheaper than traditional higher performance server computer. This also motivates us to design  our GPU based simulator. 

Rest of the paper is organized as follows:  Section \ref{prevwork} gives  the survey of previous work.  Section \ref{cachearch} present the assumed cache architecture model and management policy for our designed simulator. Section \ref{buffnet} describes about the  bufferless on chip interconnection methodology used in our simulator design. We have presented a short introduction to GPU and CUDA programming model in Section \ref{cudagpu}. Section \ref{GPUimpl} describes about both serial and parallel version implementation of simulator and its design. In  Section \ref{result}, we present our experimental result with analysis, and finally we briefly conclude the paper and discuss about future work in Section \ref{concl}.

\section{Previous work} \label{prevwork}

Performance of most commonly used shared cache (physically shared) is pathetic in terms of access latency and interference, and also as core counts of chip multicore have direct impact on  pressure on the shared on-chip cache. Many researchers proposed  technique to reduce interference either by partition efficiently or managing efficiently \cite{Qureshi,Ebrahimi201} to maximized the performance of multicore system.  

Distributed cache reduces interference and latency, but may not be good where a core requires a larger portion of cache \cite{lcmpintel}. And in that case a distributed cache slice may suffer from increased local cache pressure and eviction. So LSPD model provide an effective trade-off between cache access time and share. These papers \cite{RNUCA,victimmigration,CopCaching,ElasticCaching} describes about various variation of LSPD cache. Also they proposed many logically sharing policy, placement, eviction, replication  and management of physically distributed cache to improve the performance of multicore system. In \cite{CopCaching}, they used a mechanism called cooperative caching, which put a locally evicted block in another on-chip L2 cache that may have spare capacity, rather than evict it from the on-chip hierarchy entirely. In \cite{ElasticCaching,RNUCA}, authors proposed architecture with each core having a  cache with configurable or adoptable boundary between private and shared portion. In \cite{RNUCA}, author classified  cache access patterns of a range of server and scientific workloads into distinct classes, where each class is amenable to different block placement policies.

There are many NoC simulators and they have different features, and takes different configurable parameters. A short list of the current network on Chip simulators includes Noxim \cite{Noxim}, NIRGAM \cite{nirgam} and Nostrum  NSSE \cite{NNSE}. These simulator are cycle accurate NOC simulator developed using SystemC. In Noxim \cite{Noxim}, user can customize many parameters likes network size,  buffer  size,  packet  size  distribution,  routing  algorithm, packet injection rate, traffic time distribution,traffic pattern, and etc to simulate NoC. NIRGAM \cite{nirgam} is  a simulator similar to Noxim for NoC  routing and application modeling. Both Noxim and NIRGAM uses randomly generated traffic as input. NNSE takes topology, routing policy, and traffic patterns and based on these configuration parameters a simulator is built and executed. NNSE uses both randomly generated traffic and core generated trace traffic. These simulators mostly design to simulate network on chip target to mainly buffered network and without supporting memory hierarchy. Jantsch et. al. \cite{Eggenberger2013} explain performance different memory architecture and management in an NoC platform.  As most of the available multicore simulators like Multi2sim \cite{UJM12}, SESC \cite{sesc}, SimicGems \cite{GEMS}, and, etc are very slow.  And NoC simulators written in SystemC are extremely slow.  So we can't simulate a bigger design with more than 100 cores. Many researchers propose to enhance NoC simulation using multi-threaded parallel core or accelerated using FPGAs and these are HAsim, Protoflex and FAST \cite{HAsim,ProtoFlex,FAST}. In SimFlex \cite{SimFlex}, another group of researcher uses statistical sampling of system to speed up the simulation of targeted multicore system. DARSIM \cite{DARSIM}, HORNET \cite{HORNET1} and  Graphite \cite{Graphite} are cycle accurate parallel simulator targeted to explore multicore architecture. These simulator can be executed either in network only mode by using synthetic traces  or in full system multicore mode using a build in MIPS core simulator. These simulator target to run simulation on multicore (power full Xeon like octa or hexa core) server system. Also both the simulator uses several techniques to  speed up the simulation. These techniques are direct execution, seamless multicore and multi-machine distribution, and lax or periodic synchronization. Others parallel simulators that target multicore architecture are  COTSon \cite{COTSon}, BigSim \cite{BigSim}, FastMP \cite{FastMP} and Wisconsin Wind Tunnel (WWT) \cite{WWT}.  

	In \cite{twentyfour} authors presented first GPU-based NoC simulator without considering bufferless, that is the fastest NoC simulators built till date. To the best of our knowledge, our simulator is the first attempt on building GPU based simulator for LSPD cache system large chip multicore connected with bufferless NoC. 

\section{Cache Architecture and Management} \label{cachearch}
\subsection{Cache Architecture in Large Chip Multicore} 
\begin{figure}[tb!]
\centering%
\includegraphics[scale=0.7]{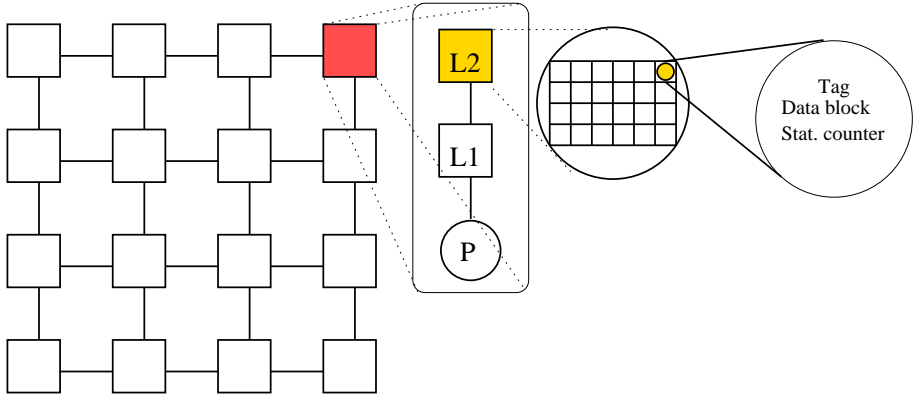}
\caption{\textbf{Large CMP with LSPD cache connected using mesh NoC}}
\label{noc}
\end{figure}

LSPD cache model on-chip L2 cache is well suited for large chip multicore. For simplicity and scalability, mesh type interconnections are popular in large scale chip multicore. Figure \ref{noc} shows an example of 16 core connected in  mesh fashion and with each core having its own local L1 cache and logically shared L2 cache slices. L1 caches are private per core due to their tight timing requirement and L1 caches are write through cache. Each core of the system have equal amount of L2 cache slices locally. The local L2 cache slice can be access very fast where as remote 
L2 slice access takes longer time. There is no replication of L2 cache block, means only one copy of L2 cache block will be present in the system at a time. 
 
As off-chip bandwidth is limited, it is essential to utilize the on-chip cache space efficiently and effectively. Also accessing a off-chip memory location is slower than accessing an on-chip remote cache location. So efficient placement, eviction, replacement and migration are very much important to manage the cache space resource in chip multicore.  

In LSPD cache model, directory contains the information of cache blocks. If there is a local L2 cache miss, then directory is consulted for on chip availability of the requested cache block in the L2 cache system. If cache block is present at some other remote core then it get  accessed from there over network. If a core is accessing a particular remote cache block very frequently then migrating  that cache block  to frequently accessing core reduces the future cache access latency for that cache block, and also reduces network  traffic. If a cache block gets migrated from a core then upon receiving a request for migrated cache block, core redirect the request packet to current location of migrated cache block till update of the migration in the directory.         

In Fig. \ref{noc}. show an extended view of cache arrangement for one core. Configuration of all core are same, each core have their own L1 cache and shared L2 cache slices. Each cache block contains tag, address, statistics counter and data block. Tag is required for identifying cache block. When a cache block is evicted then entry of the cache block must be deleted from directory. A directory is a place where we store information about all the cache blocks present in the whole on-chip L2. So, directory maintains a list of cache block and their locations.  Statistics counter maintained for keeping track of accesses from different node and it helps in migration. For example, statistics counter holds information about last $N$ (say 10) accesses.  
 
\subsection{Directory Management}

\begin{figure}[tb!]
\centering%
\includegraphics[scale=1.0]{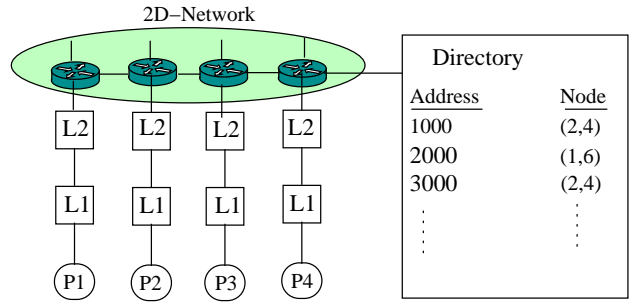}
\vspace{-0.2cm}
\caption{LSPD Cache and Directory}
\vspace{-0.3cm}
\label{lspdir}
\end{figure}
 
Fig \ref{lspdir}. shows an example of logically shared  and physically distributed cache, and it maintain a directory to keep track of location on chip cache blocks of L2 cache system. Content of directory is shown in right part of Figure \ref{lspdir}. Location of L2 cache block slice with address is (2,4), which means the cache block with tag value 100 is available on node at row 2 and column 4 of the assumed 2D mesh.  

Each node have equal size L2 cache slice that is locally accessible to the  node and also remotely accessible to all other nodes. For simplicity, we can assume a global directory. Directory maintains the location information about all  L2 cache blocks present on chip at the moment. When a node get a local L2 cache miss then the node will first consult with directory to check availability of requested cache block at other nodes. If the requested cache block is present at a remote node, then directory  will send reply with the address of node which is current owner of that block. Then a new request will made by the requesting node to local L1 cache from the remote node. Also we need to write back the evicted L1 block to the corresponding L2 slices of evicted L1 block, the location of evicted block is necessary to  write back and this location can be stored in cache block it self or directory need to be consulted.   If there is no any information about the requested block found in directory then directory will send negative reply to requested node indicating block not found. In that case requested block will make a new request to next higher level memory for that particular block.  

One can think of distributed directory, in this case directory is distributed across all the nodes. The disadvantage of centralized directory is that it may become a hot spot or bottleneck point in system. We have not considered cache coherence here. If we assume L1 is write through cache then L1 is responsible for coherence. 

\subsection{Migration and Redirection}

Clearly there are many techniques available to manage on chip last level cache data block to improve performance of LSPD cache system. There are efficient ways of placement and eviction methods for LSPD L2 cache systems. But certainly all approaches have their own advantages and disadvantages. A simple and good approach is to place the block in requester cache after that migrate to other core based on frequency of access for nodes.  Migration of a cache block may improve performance of LSPD cache system. If a node is frequently requesting to a specific cache block of a remote node, then that remote cache block should be moved to requesting node. This placement approach is local placement approach. This may lead to problem of not getting proper required shared of L2 cache space for some core. So we may need to design policy so that system should globally decide the placement, eviction and migration of the l2 cache blocks. In centralized directory organization global management is easier.

From implementation point of view we can think, within L2 cache, for each cache block have a statistics counter which contains the information about the  number of access from the remote nodes. Also it records the number of access from all remote nodes in a list for last $N$ accesses. If a remote node have accessed it mostly, then migration get triggered.  If a local node have accessed it most time than there is no need of migration. If migration is required, whole cache block will be sent/migrate to that remote node. While migration of cache blocks is in transit, source node will continue to provide service for migrated cache block till migrated cache block reached its destination. When migrated cache block reaches at its destination, source packet invalidates its copy and directory get updated.  A statistics counter need to be placed in every cache block to implement the migration. The check for a local cache migration may get triggered when a request comes from remote nodes. 

If a node is servicing a request for a particular L2 cache block then that block will not get migrated till current request is serviced. If a node, say $n_1$,  make a request for particular L2 cache block at time $t$ (say) to node $n_2$. while request is in transit, in mean time if desired cache block gets migrated from node $n_2$ to $n_3$. Then upon receiving the request from node $n_1$, node $n_2$ will make a new request on behalf of node $n_1$. It will redirect the request by changing destination address, since directory has information of all L2 cache block. It prevents from dropping of request cache block. 

\subsection{Cache Block Replacement Events}
We have assumed local victim selection for replacement. If a chosen victim cache block position in directory already contains a valid cache block then that cache block entry is deleted since it is no longer accessible after replacement. Replacement routine gets triggered at the following time: 
\begin{itemize}
\item When a new cache block comes to L2 cache from main memory.
\item When a migrated cache block from remote node comes in to a local L2 cache.
\item A L1 cache block replacement happens when a new block comes in to local L1 cache either from local L2 cache or from remote L2 cache.
\end{itemize}
The evicted block need to be written back to the memory or corresponding L2 block if it is a L1 block replacement. 
\subsection{Request Policy}
When a node make a request for remote L2 cache block in network then it will not send another request packet till it receives the cache block  from remote node or from memory. In the mean time it can serve to all local requests. That means miss under a miss is not allowed, but a hit under a miss is allowed. As in bufferless network data flits may arrive in out of order, it is necessary to keep a re-order buffer to order the flits of a cache request/access to make a complete cache block or request.

\subsection{Interface Buffers with Network System}
In our to be simulated target architecture, we assume each node has receiving buffers of L2 block size. Data from the network get received and is stored in receiving buffer.  These data may be  L2 cache block or L1 cache block or any other control messages. After completion of data reception for a particular event, the data block from the buffer is transferred to computer system (or written to L1 or L2 cache). Similarly when a node sends or migrates a block to other node through network, it placed the data block into sending buffer before starts sending. Since for a request/cache block transfer many flits are send,  and each flit routed independently. So these flits may follow different paths and  take different travel time to reach its destination. And these flits may reach in out of order at receiving node.  Receiving node will put all the  flits in buffer according to its number. When all flits of a access or cache block migration arrived at destination then it get delivered to local core of destination router \cite{lcmpintel}. Table \ref{FF} shows typical flits division of different request, access and migration of cache blocks.
\begin{table}[tb]
\centering 
\begin{tabular}{|l|c|c||l|c|c|}	 
	\hline
\textbf{Task } & \textbf{Type} & \textbf{flits}  & \textbf{Task } & \textbf{Type} & \textbf{flits} \\  [0.5ex] 	\hline
Directory access  & DA & 1 & L2 Blk Replacement & B2 & 16\\ 	\hline
Directory reply  & DR & 1 & L2 Blk Migration & B2 & 16 \\ 	\hline
Reply redirection  & RR & 1 & Remote L2 access & RA & 4\\ 	\hline
\end{tabular}
\caption{Flit division of request/access}
\label{FF}
\end{table}

\section{Bufferless Network on Chip} \label{buffnet}

\subsection{Need of Bufferless NoC}
Minimizing power consumption and decreasing latency are critical design challenge in implementing interconnection network of large chip multicore. Also as core count in modern CMPs continues to increase, the interconnection network becomes a more significant component of system hardware cost.  Several NoC prototype points toward this trend. And these are (a) interconnect consumes 40\% of system power in MIT RAW chip \cite{twentyeight} and (b) 30\% in Intel Terascale chip \cite{twentyseven}, etc. Buffers consume a significant portion of this interconnect power. Removing or minimizing in-router buffer are alternate designs to solve this power consumption challenge. Bufferless NoC design  evaluated by Multu et. al. \cite{fifteen,two}, suggest as an alternative to traditional virtual-channel buffered routers. The bufferless interconnection design appealing mainly for two reasons: reduced hardware cost, and simplicity in design. Also they have shown in \cite{two}, by eliminating buffers in network provides an energy reduction of 40\% with minimal performance impact at low-to-medium network load.  

\subsection{Division of Packet into Flits and Routing Flits}
Each packet (or data  or cache block) gets fragmented in a small routable unit called flit. Each flit of a data or cache block is routed independently. For this each flit contain routing information or header. In bufferless interconnection, each router has equal number of incoming and outgoing links and this number is 4 for mesh interconnection. Flits that are contending for a common port, the lower priority flit gets deflected to a free port. As there are four in coming ports and four out going ports in each router, we can ensure that there exist a free port to deflect low priority flit of the contending flits. A node itself can inject a packet into network whenever it finds a free input port. Received flits may arrive in out-of order, so we need reorder buffer at destination. When all flits of a packet (or data or cache block) is received then it is delivered to receiving the core.  

If we use age based priority, flit will always reach at its destination \cite{one}. No livelock will occur because once a flit is the oldest flit in the network, it get highest priority and chance of getting deflected is zero and this guaranteed to make forward progress until it reaches its destination. Eventually, a flit will become the oldest flit in the network, after which it cannot be deflected any more. In our case, we have chosen fully adaptive XY-routing algorithm also known as prioritized multi-dimensional routing (PMDR) \cite{seven} to for our  simulator. If a flit does not get deflected than it will first travel in x-direction, making x-offset zero then it will make y-offset zero. If a flit gets deflected then still it tries to travel first in x-direction and then in y-direction. When a flit get deflected its age get incremented.

\begin{figure}[tb!]
\centering%
\includegraphics[scale=0.75]{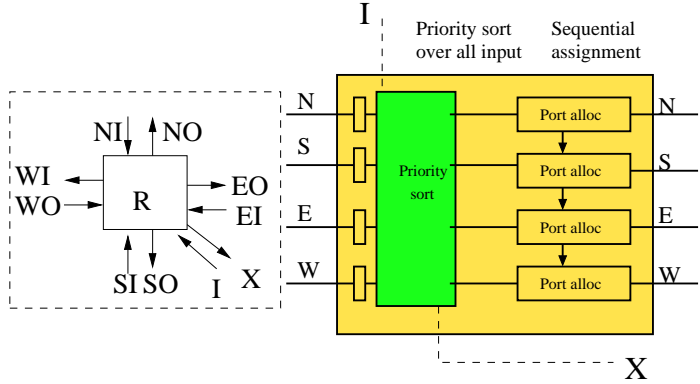}
\caption{\textbf{Bufferless router}}
\label{router}
\end{figure}

\subsection{Bufferless Router} 

Bufferless router does not  have buffer in router to store the packets. To implement a bufferless router in hardware, we require 4 registers, a crossbar, and arbitration logic. Registers and crossbar are used for storing and transferring of flits within router. The work of arbitration logic is  to assigns output port to incoming flits. It consists of two components and these are (a) ranking component and (b)  port-selection component. All the incoming flits are to sorted based on their ages (older flit get higher priority)  by ranking component which is `Priority Sort' block as shown in Figure \ref{router}.  	
  
The router considers the flits one by one in the order of their age (highest age first) and assigns to each flit the output port with highest priority that has not yet been assigned to any higher aged flits \cite{three}.  As shown in Figure \ref{router}, there are two extra links present in the bufferless router which are connected to the associated core. One is inject link (I) and other is eject link (X). When a core wants to inject flits into network then it place flits to free input ports upon availability of free input port of the router associated with the core. If a flit is destined for a particular core then it is ejected by router through eject link of  the associated core with the router.  

\section{GPU Architecture and CUDA Programming} \label{cudagpu}

GPUs are designed to maximize the throughput of instructions by executing as many instructions as possible at the same time. CPU design dedicates small amount of chip area for compute (example 4 ALU's) and higher area for cache and complicated control mechanism. Whereas GPU dedicates significantly higher amount of area for compute part rather than cache and control \cite{seventeen}. As a result,  number of ALU in a GPU is much higher than of a CPU. So, GPU can be used for parallel parts of the program where a lot of operations ( $\ge 10^4$) that can be done in parallel. GPUs consists of an array of Streaming Multi-cores (SMs). Each SM has a set of tiny cores or streaming cores (SPs also known as CUDA core) along with a small amount of common shared memory. A small part of code that is executed on GPU is called kernel and executes huge number of times  on independent data  and that can be run parallel in single program multiple data (SPMD) model. In CUDA programming model, a pool of parallel thread can be hierarchically organized as block of threads and grid of blocks. When a kernel is launched, a grid of blocks (consist of threads) is executed on GPU. GPU thread scheduler map a block of threads to one SM where they share a common memory and all the threads within block can communicate with each other with the help of shared memory. In our work, we have used NVIDIA GeForce GTX 690 GPU for our experimentation. Specification of used GPU is shown in Table \ref{spec_gtx690}.
\begin{table}[tb] 
\centering
\begin{tabular}{ |l|l| }\hline
  \textbf{GPU config} &  \textbf{Amount or count} \\
	\hline
Number of SM  &  16 \\
	\hline
Number of SP per SM & 192\\
	\hline
Shared memory per block & 49,152 bytes \\
	\hline
Maximum threads per block & 1024 \\
	\hline
Total constant memory per device & 65,535 bytes \\
	\hline
Maximum no. of threads per SM & 2048 \\
	\hline 	
\end{tabular}
\caption{Specification of Nvidia GeForce GTX 690} 
\label{spec_gtx690}
\end{table}

\section{Implementation Simulator Modules } \label{GPUimpl}

\subsection{Design Strategies}
In this Section, we have described implementation of our designed simulator to simulate on chip cache management of large chip multiprocessor where cores are connected using bufferless mesh interconnection. Clearly this work is divided in two parts: (a) serial version of simulator, and (b) parallel GPU version of simulator using CUDA programming model. First we have implemented serial version (using C++) of simulator, and assumed this as golden model, and used this for correctness verification of parallel version of simulator(implemented using CUDA). 

The designed simulator consists of many building blocks or modules. These modules are (a) bufferless deflection routing module (b) LSPD L2 cache  with directory and migration module, and (c) trace generation module. These modules or building blocks are described in next subsection.
\subsection{Building Blocks of Simulators}
\subsubsection{Bufferless Deflection Routing Module}
In our case, we have assumed the cores of LCMP connected using bufferless routers organized in two dimensional mesh fashion. So our assumed network system consists of two dimensional mesh of bufferless router and each router associated with a compute core. We can simply call the router and it's associated core as node. 


Each router has four incoming ports and four outgoing ports. In case of virtual channel (non bufferless) routers have buffer at incoming ports (or outgoing ports). But in bufferless network, network system do not  have buffer means router do not  have buffer, but core attached to the router have sending space/buffer and receiving space/buffer.

We have defined the structure of flit, router and core as given bellows: 
\small
\begin{lstlisting}
struct Flit{
 int	FlitId, PacketId,FlitSize, Age;
 int	TotFlitsOfPacket,SrcX,SrcY,DstX,DstY;
 char	*FlitData;
};
struct Router { 
 Flit	Input[4], Outport[4], InFromProc, XToProc;   
 int	Direction[4];
 bool	InActive[4],OutPSignal[4],InActive,XActive; 
 struct core *AssoProc;
} 
struct Core {
 Flit		ToBeSend[MaxFlitsPerData];
 int		NextFlitAddress;
 Flit		ReOrderBuffer[MaxFlitsPerData];
 bool		Wait;
 LSPDSlice	L2;
 Cache		L1;
 long int	ReplySent, Req, Reply, ReceivedReply; 
 long int	ReqReceived, Redir, Trap;
} 
\end{lstlisting}
\normalsize
Structure `Flit' is defined to hold information about flit. It contains source address, destination address, flit id, packet id, total number of flits of the packet, age, size of pay load data of flit and payload data of the flit. Additionally flit has one age field which is used while routing flit and this age field incremented by one each time when it get deflected at a router. 

Structure router have Input[0], Input[1], Input[2] and Input[3] to  hold input flit (if available) coming from north, south, east and west direction of the router respectively. `InActive' array is used to check the availability of flit at their input ports. Array `Direction' holds the desired outgoing direction of the incoming flits.`OutPort' array holds the outgoing flit which is assigned by router using bufferless deflection routing algorithm. Also direction need to be calculated based on the destination address. Array `OutPSignal' signals the outgoing flit at their corresponding port. It helps in transferring of the flits between neighbor routers.

Structure `Core' is the structure definition of a core  to hold many useful information related to processor (or core) activity. Array  'ReOrderBuffer' receive flits and array `ToBeSendBuffer' holds any outstanding flits to be sent. Boolean variable `wait' is set when a core waits for reply. Since packet/data gets fragmented in to flits, so `ReOrderBuffer' hold the receive flits and the associated processor compile receive request after receiving all the flits of the same request. For statistics generation, there are several variables present inside structure of core are like number of request, reply, trap, redirection, etc. Trap is situation in which, a router received a request for cache block but requested cache block is neither present at this router nor directory have any information about this block. In this case it will send the invalid packet to requested core. So in that case data need to bring from memory. 

When flits arrived at router, sorting of flit is done according to their age. Oldest flit have highest priority and will not be deflected. After sorting, flits get assigned to their outgoing port. If some flits are contending for same direction, then oldest flit will win and rest of the flits are deflected to the available free ports. When a flit get deflected from one router to another router its age get incremented. This age increment prevent from live lock and deadlock situation. 

Core associated with the router is responsible for breaking the to be send data  in to flits and put on to ToBeSendBuffer. Also when all the flits a request/packet are arrived to the core ReOrderBuffer, core re-assemble all the flits to make a complete packet/request data and use that data. 

\subsubsection{Simulating LSPD L2 Cache with Redirection and Migration}


\begin{figure}[tb!]
\centering%
\includegraphics[scale=0.6]{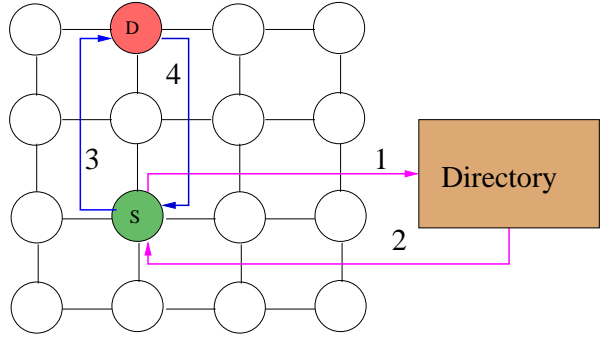}
\caption{\textbf{Sequence of events occurs in a memory access}}
\label{memory}
\end{figure}

As explained earlier each router is associated with local core which has a private L1 cache and some L2 slices of total LSPD L2 cache system. Since this L2 cache slice is LSPD cache, all L2 cache block is accessible to all other cores. Only one copy of a specific L2 cache block/data will be present at any time in the system. 

Address generated by core get checked in local L1 cache and if it is miss then it is checked in local L2 cache slices. As shown in Figure \ref{memory}, if there is a miss in local L2 slices of the core `S' then global directory is consulted (in Fig. \ref{memory} annotated with 1). If required cache block is present on chip at another remote node then directory will send information about that remote core in reply (in Fig. \ref{memory} annotated as 2). After receiving a positive reply from directory, the source core sends a request to the remote core `D' which currently holds the required cache block (in Fig. \ref{memory} annotated as 3). And finally, if the requested cache block is present at the remote node then that remote L2 cache slice will be access and data get copied to L1 cache of local core L1 cache `S' (and this is annotated as  4 in Fig \ref{memory}). If the request cache block is not present on the chip then directory will send negative acknowledgment to requested core `S' and in this case block will be fetched from main memory and directory will be updated.  Remote cache slices of core `D' is accessed means it copy a data of size L1 cache block size to local L1 and put into local L1 cache of source core `S'. When this block get evicted, then this needed to be written back to the corresponding destination L2 cache slices. 

In our case, we have  used centralized directory, but centralized directory may be a bottle neck to access the directory. Assume the centralized directory designed implemented using a associative memory or hardware parallel dictionary, so that it can be accessed in constant time. But in simulator, it is difficult to use associate memory. Instead, in our simulator we used a very funny and puny trick to handle this issue and that is by  using a large location array containing (x,y) position of the cache block indexed by tag of address. So for example if our to be simulated targeted multicore architecture  has 1GB of total memory with 64B of L2 cache block then size of location array will be $\frac{2^{30}B}{64B}=2^{24}.(2B+2B)=16MB$ by assuming 16 bit signed representation of x and y. In actual  associative memory hardware implementation, amount of memory required to store the directory will be of size 4MB assuming 64 MB of on chip LSPD cache with 64B L2 line size. Distribute directory may improve the network performance as remove the hot spot, but searching and updating distributed directory is a critical issue. 

Our simulator can generate various simulation  statistics like total request made, total received request, total reply send, total reply received, total migration, total number of redirection, total cache hit and cache miss for each cache, total directory access, total trap, and, etc..
\begin{figure}[tb!]
\centering%
\includegraphics[scale=0.65]{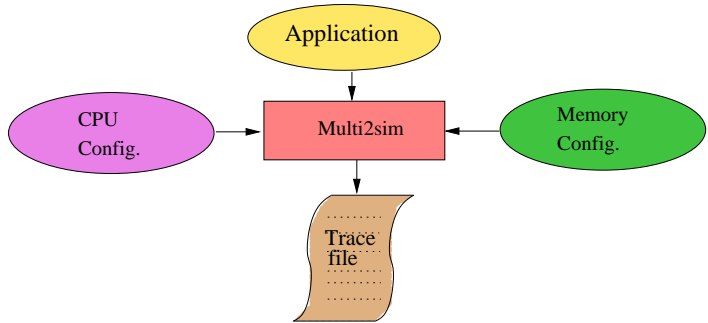}
\caption{\textbf{Trace generation}}
\label{trace}
\end{figure}

\subsubsection{Generating Trace}

Input to our simulator can be either a trace driven injectors, or random traffic generator, or a cycle accurate MIPs like core. We are taking  memory access address trace as input to the LSDP cache system of on chip multicore connected using bufferless router.  In trace driven simulation, trace provides the memory access addresses to core and core reads addresses from this trace file and use for simulation. For generating representative trace, we have used Multi2sim simulator \cite {UJM12}. Multi2sim simulator requires CPU configuration file and memory configuration file along with GCC compiled x86 executable of application to generate trace as shown in Figure \ref{trace}. We have used 64 core CPU configuration with memory configuration fixed with various parameters of L1 and L2 cache like associativity, block size, latency etc. We have generated the "representative traces" corresponds to five different applications to test our simulator. In trace, we are capturing the addresses generated by cores. The trace used is representative trace and we can't generate access trace for more than 100 core using  Multi2sim like multicore simulator. Some time these type of simulator may not support more than 100 core or it may take huge amount of time to generate trace. 

\section{Overall serial and parallel version of Simulator}

\subsection{Serial Version} \label{serversion}
In general, serial version of our simulator consists of a main loop performing three specific procedures repeated till simulation is not finished. A simplified form of this simulator is as follows: 
\small
\begin{lstlisting}
SimCycle = 0;
while(SimulationNotFinished) {
	CheckForSimulationFinished();
	PutAccessRequestsToAllCores();
	for(i=0; i<RouterCount; i++) Phase1(i);				
	for(i=0; i<RouterCount; i++) Phase2(i);
	for(i=0; i<RouterCount; i++) Phase3(i);
	GetStatusOfAccReqForAllCores();
	SimCycle++; if(SimCycle>MAXSIMCYCLE) break; 
} 	
GetSimulationStatics();	
\end{lstlisting}
\normalsize

Works carried out in all these three phases are described bellow:
\subsubsection{Work in phase one}
 In this phase, we simulate  memory access in LSPD cache. This work need to be done by all the core/node. If there is no pending memory access request of own core then the core access memory specified in next address to be accessed in his own next address access list. The current access of the core is a hit in local L1 cache, it serve immediately and finish work of the phase for the cycle. Other wise it will wait up to L1 miss cycle, means it will mark there is a pending request and set a counter to L1 miss cycle and finish the work of phase till it reaches the counter to 1. This means for next some cycles (L1 cache miss cycles) there will be no simulation work for this core in this phase. Next time when it enter to this phase, the core check for pending request and decrements the counter for the current cycle, if counter reaches 1, then it access local L2 cache slices of for the missed L1 block. If there is a  hit in local L2 slice, first a victim cache block from local L1 cache will be selected and write back the victim back to location of victim block and the accessed block from local will be moved to local L1 cache.  Other wise need to get cache block from remote one and also write back the victim back to location of victim block. The procedure of get cache block from remote one and also write back the victim back to location of victim block need to use the bufferless interconnection network. Sending cache block data by fragmenting into flits to be done on third phase. In most of the cycle, in this phase, the core will simply increment a counter or check a wait status.       
\subsubsection{Work in phase two} 
In phase two, we have implemented the buffer less deflection routing. In this phase, each router sort the all incoming flits including injection flits according to age and assign output port according to their priority. If there is a conflict it simply deflect to any free port except the ejection port.  Newly injected flits age is set to zero.   
\subsubsection{Work in phase three}
In this  phase, each router transfer flits to next router (or get/collect flits from outputs of neighbor routers to inputs of self), increment the ages of flits (if it get delfected) and assign direction in flits according to XY routing assuming current node as source for all the flits. It also perform the work of flit ejection/receiving, sending flit to network system through injection port, dividing data into flits and re-order the incoming flits. After reordering flits it modify the waiting status to zero, which indicated that request reply arrived.

\subsection{GPU Version}
Corresponding to CPU simulation, the pseudo code of GPU implementation of simulator is given as follows:
\small
\begin{lstlisting}
int TrcHost[ROW*COL],*TrcDev,SimCycle=0;;	 
double size;
size = ROW*COL*sizeof(int);	 
cudaMalloc((void **)&TrcDev,size);	
dim3 DGrd(BlkXDim,BlkYDim,BlkZDim), DBlk(ThrdXDim,ThrdYDim,ThrdZDim);
initialize <<<DimGrd,DimBlk>>>(ROW, COL);  
while(simulation_not_finished){ 		  		 
 loadInput(TrcHost);	 		 
 cudaMemcpy(TrcDev,TrcHost,size,CopyFlag);		
 LSPDSimFromTrce<<<DGrd,DBlk>>>(ROW,COL); Synch();
 FltsPrtAsgnOrDef<<<DGrd,DBlk>>>(ROW,COL); Synch()		
 TransRoutAgeCal <<<DGrd,DBlk>>>(ROW,COL); Synch()
 SimCycle++;
 if(SimCycle>MAXSIMCYCLE) break;
 GetStatusOfAccReqForAllCores();
}
 GetSimulationStatics();	
\end{lstlisting}
\normalsize
Initially we are allocating memory on GPU for input data (trace). Before launching the kernel, dimensions of grid and block of threads need to be defined. All the kernel functions, we are launching with number of threads equal to the number of routers or cores in the target system to be simulated. Work of each router is assigned to one CUDA thread. CPU read data from trace and transfer on to GPU because at the run time GPU thread can't access memory of host computer. As size of traces are very large (in GBs), we can not store in GPU memory.   

At the beginning of the program execution whole router and caches are initialized, and it is accomplished by launching initialization kernel. When a kernel is launched then all the threads inside that kernel execute same program code. By calling 'Synch()' or  `cudaThreadSynchronize()', all threads synchronize with each other. If a thread completed it's task, then it will wait for all other threads to complete. When all the threads inside a kernel finished executing his task then kernel terminates. Main loop of the simulator is a while loop, which consist of three kernels. These  kernels are LSPDSimFromTrce, PrtAsgnOrDef and TransRoutAgeCal. Work to be done in kernel functions LSPDSimFromTrce, PrtAsgnOrDef and TransRoutAgeCal are same as work done in Phase 1, Phase 2 and Phase 3 of serial version of simulator respectively and it described  in Section \ref{serversion}. All the above three kernels repeats until core generated memory request finishes and all flits gets transferred (means there should be no any outstanding flit in the network).  
\subsection{Validation}
We have verified the result of  GPU based parallel simulator with the result of serial simulator. For same trace, the results are same for both the simulators. For verification, we have simulated up to 2000 core, in both the serial and GPU based parallel simulator and verified the result of correctness. Furthermore, we have simulated up to 43,000 cores on GPU based parallel  simulator without validating with the  serial version. 

\section{Experiment and Result} \label{result}

We have tested our simulator with representative traces of different multi-threaded benchmarks and  SPEC OMP 2001 benchmarks. Benchmarks  used to test our simulator are `matmul', `wupwise', `mgrid', `equake' and `apsi'. Partial results of GPU based parallel simulator for these application traces are shown in Table \ref{traceResult}. In that table `Request made' column displays the total number of requests  made to the remote node and it does not  count the request made for main memory. These results are obtained using 10,000 cores configuration using our GPU based  simulator. We have shown some important statistics in the Table \ref{traceResult}. Apart from these statistics simulator also displays total L1 and L2 cache hit and miss for each cores of simulated architecture. Additionally it shows the total number of migrations which occurred during simulation. `Trap' is a case in which a core receives a request for a cache block but could not processed it because of unavailability of that cache block at the remote node in the system. 

\begin{table}[tb!] 
\centering
\setlength{\tabcolsep}{0.8mm}
\begin{tabular}{ |l|r|r|r|r|r|r|r|r|}	 
	\hline
\textbf{Appl} & \textbf{Req.} & \textbf{Req.} & \textbf{Reply}  & \textbf{Reply}  & \textbf{Trap} & \textbf{Redir-} & \textbf{Dir-} & \textbf{Mem}\\  [0.5ex]
 &  \textbf{made}  &  \textbf{Rcved} & \textbf{sent} &  \textbf{Rcvd} &  & \textbf{ection} & \textbf{Srch} & \textbf{Req}\\
	\hline
\textbf{matmul} & 15,879 & 15,879 & 44,344 & 44,344 & 4,729 & 64 & 44,698 & 28,739\\
	\hline
\textbf{apsi}  & 9,450 & 9,450 & 31,308 & 31,308 & 1,567 & 56 & 24,273& 12,734 \\
	\hline 	
\textbf{mgrid}  & 14,785 & 14,785 & 55,332 & 55,332 & 932 & 20 & 33,280 & 15,232 \\ 
	\hline
\textbf{wupwise}  & 13,675 & 13,675 & 40,644 & 40,644 & 3,455 & 59 & 48,395 &  32,092 \\
	\hline
\textbf{equake}  & 16,390 & 16,390 & 54,064 & 54,064 & 2784 & 90  & 42,926 & 23,836  \\
	\hline

\end{tabular}
\caption{Simulation Result of Different Application Traces} 
\label{traceResult}
\end{table}

\begin{figure}[!tb]
\centering
\includegraphics[scale=0.8]{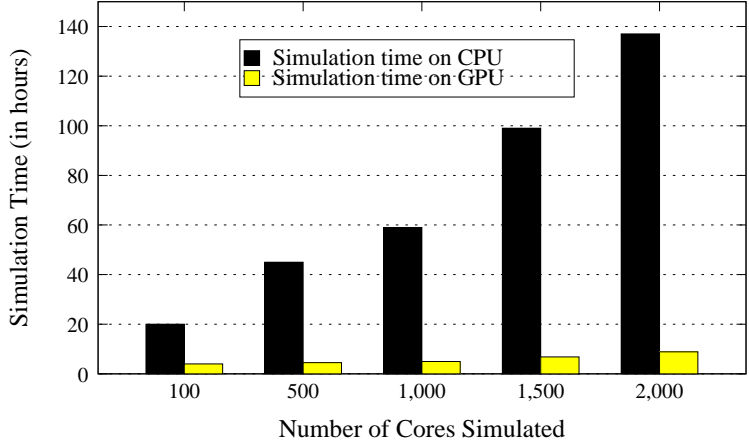}
\caption{\textbf{Comparison of simulation time of serial version vs parallel GPU version}}
\label{speedup}
\end{figure}

We have simulated the serial version and GPU based parallel simulator for different number of cores and measured speedup. The results of simulation on `matmul' application trace for 100, 500, 1000, 1500 and 2000 cores is shown in Figure \ref{speedup}.  Input trace size is proportional to the number of core of to be simulated architecture. In our case, the input trace for N core system simulation is traces with $N \times M$ (say M=200) address references. So for 100, 500 and 1000 core system, trace sizes are $100 \time 200$, $500 \times 200$ and $1000 \times 200$ respectively. As  shown in Figure \ref{speedup}, simulation execution time of serial version increases rapidly with the  number of cores while parallel GPU version simulation execution time is much faster as compared to serial one. We have achieved a speedup of 25 in simulating a targeted 2,000 core system. 

As Figure \ref{speedup} suggests, for higher number of core, execution time of serial version of simulator is increasing rapidly with number of cores in to be simulated system. We were not able to simulate serial simulator for more than 2,000 core. In GPU simulator we have simulated up to 43,000 cores. This is the upper limit of number of core for our GPU based parallel simulator. The limitation of 43000 CUDA threads is due to amount of available memory on GPU Nvidia GeForce GTX 690 GPU. At different cache configuration, maximum number of simulated core has been shown in Table \ref{cacheproc}.

\begin{table}[tb!] 
\centering
\begin{tabular}{|l|l|r|}	 
	\hline
\textbf{L1: Set, Assoc}      & \textbf{L2: Set,Assoc} & \textbf{Max. simulated}  \\  [0.5ex]
 \textbf{ \& Line-size}   &  \textbf{\& Line size}  & \textbf{cores} \\  [0.5ex]
	\hline
128,4,32 & 512,8,64 & 2,000\\
	\hline
128,4,32 & 128,4,64 & 10,000\\
	\hline
32,2,32 & 32,2,32 & (with migration) 30,000 \\ \hline
32,2,32 & 32,2,32 & (without migration) 43,000 \\ \hline  
\end{tabular}
\caption{Cache configuration and maximum simulated core} 
\label{cacheproc}
\end{table}

In our simulator, we store data for all the routers, cores and caches on the global memory of GPU. Global memory is limited in size. Data structure to simulate cache takes significant amount of memory of GPU. So with the varying of cache size, more number of threads can be executed. In the Table \ref{cacheproc}, the upper limit of threads with different cache size has been shown. Also implementation of migration requires some amount of memory for each cache blocks. If to be simulated system  is large, where huge number of cache blocks are present then migration implementation consumes significant amount of memory to store access history of each cache blocks. So by removing migration strategy, we were able simulate more number of cores. In the third row of the Table \ref{cacheproc}, for given cache configuration, we were able to run 30,000 core with migration and 43,000 core without migration.

\section{Conclusions} \label{concl}

Simulation of large scale CMP with on chip interconnection have huge scope of parallelism and we have exploited this parallelism using a modern GPU to achieve significant speedup as compared to serial execution of our designed simulator. We have achieved 25 times speedup in the trace based simulation. By simulating the more cores in to be targeted system, we can achieve more speedup in our GPU based parallel simulator. This simulator can be used for various scientific and research purpose. 

We have used centralized directory and also local cache replacement policy, which will not give good performance in reality. By using global cache replacement policy using centralized directory, performance can be improved of the targeted system.  In future, we are planning to  implement distributed directory for more realistic case. It will improve the performance and hence speedup. As we have not considered cache coherence in our simulator, it can be very good extension. Also, modeling global placement, migration and eviction in simulator may add substantial value to simulator.  

In future, we are planing to extended our simulator to support MIPS like core to simulate real applications instead of trace driven traffic or synthetic generate traffic based simulation.

\end{document}